\documentclass{article}
 \usepackage{graphicx}
\usepackage[cp1251]{inputenc}
\begin{document}

 \centerline  {\large\bf Kinematics of pre-main sequence stars from the Gaia DR2 catalog}
\bigskip
 \centerline   {V. V. Bobylev\footnote{E-mail: vbobylev@gaoran.ru} and A. T. Bajkova}
\medskip
 \centerline   {\small\it Central (Pulkovo) Astronomical Observatory of RAS, 65/1,}
 \centerline   {\small\it Pulkovskoye Chaussee, Saint Petersburg, 196140, Russia}
\bigskip
\bigskip
 \centerline   {ABSTRACT}
\medskip
The kinematic properties of young stars that have not yet reached the stage of the main sequence are studied. The selection of these stars was recently carried out by Marton et al. (2019) and Vioque et al. (2020) according to the Gaia\,DR2 catalog using a number of photometric infrared surveys. We have determined the rotation parameters of the Galaxy and the parameters of the ellipsoids of the residual velocities. The linear velocity of the circular rotation of a solar region around the center of the Galaxy, found using 4431 stars, is equal to $V_0=229.1\pm4.4$~km s$^{-1}$. The following ellipsoid parameters of their residual velocities are found from low-mass stars (T\,Tau type stars): $\sigma_{1,2,3}=(9.45,6.99,6.61)\pm(0.94,0.43,0.32)$ ~km s$^{-1}$. For stars of intermediate masses (Herbig Ae/Be stars), their values turned out to be somewhat larger $\sigma_{1,2,3}=(13.67,9.25,7.26)\pm(2.40,2.44,0.88)$ ~km s$^{-1}$. Distant stars from both Catalogs trace the local spiral arm well. For 1212 stars, a new estimate of the pitch angle of the Local spiral arm is equal to $i=-8.9\pm0.1^\circ$.


 \newpage
\section{INTRODUCTION}
 Recently, stars that have not reached the main sequence stage (PMS) have attracted great interest of researchers. These include both Herbig Ae/Be stars (HAeBes) stars with masses from $2M_\odot$ to $10M_\odot$, and stars like T\,Tau with masses less than $2M_\odot$. In studying the structure and kinematics of the Galaxy, such stars are important due to their exceptional youth. However, until recently, astrometric data necessary for analysis were known only for several hundreds of such stars located in the Gould Belt region, i.e. near the Sun (Wichmann et al., 1998, Bertout et al., 1999, Mamajek et al., 2002). Nevertheless, their analysis led to a deep understanding of the evolution of the stellar associations closest to the Sun (Preibisch \& Zinnecker, 1999). The example of the Sco-Cen association showed, in particular, that there are no significant differences in the kinematics of massive OB stars and less massive PMS stars~(Sartori et al., 2003).

 With the advent of the Gaia\,DR2 (Gaia Collaboration 2016, 2018a) catalog, it became possible to study a large number of such stars. The catalog contains trigonometric parallaxes and proper motions of about 1.3 billion stars. For a relatively small fraction of these stars, also their line-of-sight velocities were measured. Photometric measurements are presented in two broad bands, so only a very rough classification of stars is possible. For reliable classification, it is necessary to use more accurate spectral and photometric data from other sources. Nevertheless, according to the Gaia DR2 catalog, a number of important studies have been carried out related to the kinematics of various Galactic subsystems. For this, as distant young objects, for example, OB stars~(Xu et al., 2018, Bobylev \& Bajkova, 2018, classical Cepheids~(Mr\'oz et al., 2019) or open star clusters~(Soubiran et al., 2018, Bobylev \& Bajkova, 2019a) were used.

 More than 11\,000 high-confidence pre-main sequence members of Sco\,OB2 were selected from the Gaia\,DR2 catalog in work of Damiani et al. (2019). More than 11\,000 Young Stellar Objects (YSO) are highlighted in the Vela OB association area by Cantat-Gaudin et al. (2019). The structure of the association in the Orion arm was studied using 1000 YSO with data from the Gaia\,DR2 catalog in work of Grosschedl at al. (2018). In the region of the Gould Belt, i.e. in a near-solar neighborhood with a radius of 500 pc, Zari et al. (2018) selected more than 40\,000 stars of the T\,Tau type according to kinematic and photometric data.

 A list of over 1.1 million YSO candidates has been compiled in work of Marton et al. (2019), which are already distributed in a wider neighborhood of the Sun. Using the YSO sample from this catalog~(Krisanova et al., 2020), obtained estimates of the rotation parameters of the Galaxy. Their values are in good agreement with the estimates of other authors obtained from other young objects. However, we were faced with the fact that the stars we selected had a rather large dispersion of residual velocities, about 16~km s$^{-1}$, which is higher than the expected value for young stars of $\sim$ 10~km s$^{-1}$. Apparently, the sample was contaminated with older stars. In addition, the sample contained stars with relative trigonometric parallax errors of less than 20\%. In this paper, we want to use another division of stars into groups, and also consider stars with parallax errors of less than 10\%.
 Vioque et al. (2020) compiled a completely new catalog of Herbig Ae/Be, classical Be and PMS stars. It contains more than 10,000 stars with kinematic and photometric data from the Gaia DR2 catalog. The selection of these stars was carried out according to a method different from that used in Marton et al. (2019).

 The aim of this work is to compile PMS stars samples from the works Vioque et al. (2020) and Marton et al. (2019) with the least amount of contamination and to determine from them the parameters of the rotation of the Galaxy and the parameters of the ellipsoid of the residual velocities of stars.

 \section{Method}\label{method}
\subsection{The Galaxy rotation curve parameters}
We use a rectangular coordinate system centered on the Sun, in which the $x$ axis is directed toward the Galactic center, the $y$~axis is toward the Galactic rotation, and the $z$~axis is toward the north pole of the Galaxy. Then $x=r\cos l\cos b,$ $y=r\sin l\cos b$ and $z=r\sin b.$
From observations there are known three components of a star velocities: the line-of sight velocity $V_r$ and two projections of the tangential velocity $V_l=4.74r \mu_l\cos b$ and $V_b=4.74r\mu_b$, directed along the Galactic longitude $l$ and latitude $b$ respectively. All the velocity components are measured in km s$^{-1}$. Here, the coefficient 4.74 is the ratio of the number of kilometers in an astronomical unit to the number of seconds in a tropical year, and $r=\pi^{-1}$~is the heliocentric distance of the star in kpc, which we calculate through the parallax of the star $\pi$ in mas. The components of the proper motion $\mu_l\cos b$ and $\mu_b$ are expressed in mas year$^{-1}$.

To determine the parameters of the Galactic rotation curve, we use equations obtained from the Bottlinger formulas, in which the angular velocity $\Omega$ is expanded into a Taylor series in powers of $(R-R_0)$ to terms of the second order of smallness $r/R_0$:
 \begin{equation}
 \begin{array}{lll}
 V_l= U_\odot\sin l-V_\odot\cos l-r\Omega_0\cos b\\
 +(R-R_0)(R_0\cos l-r\cos b)\Omega'_0
 +0.5(R-R_0)^2(R_0\cos l-r\cos b)\Omega''_0,
 \label{EQ-2}
 \end{array}
 \end{equation}
 \begin{equation}
 \begin{array}{lll}
 V_b=U_\odot\cos l\sin b + V_\odot\sin l \sin b-W_\odot\cos b\\
    -R_0(R-R_0)\sin l\sin b\Omega'_0
    -0.5R_0(R-R_0)^2\sin l\sin b\Omega''_0,
 \label{EQ-3}
 \end{array}
 \end{equation}
where $\Omega_0$ is the angular velocity of the Galaxy at a solar distance $R_0$, $\Omega_0^{(i)}$~is the $i$-th  derivative of the angular velocity with respect to $R$, the linear rotation velocity at a solar distance equals to $V_0=R_0\Omega_0$, $R_0$~is the galactocentric distance of the Sun, $R$~is the distance from the star to the axis of Galactic rotation, $R^2=r^2\cos^2 b-2R_0 r\cos b\cos l+R^2_0.$

The well-known Oort-Lindblad kinematic model (Ogorodnikov, 1965), where two expansions are performed in a series -- the expansion of the angular velocity $\Omega$ in powers of $R-R_0$ and the expansion of the distance $R$ in powers of $r$. As a result, the applicability of the Oort-Lindblad model is limited by the distance $r\simeq2$~kpc. The range of applicability of the model (\ref{EQ-2})--(\ref{EQ-3}) is much wider, 5--6 kpc. In addition, this area can be easily expanded by adding more and more expansion terms. The use of two terms of the expansion, as in our case, is good because it allows one to obtain the rotation curve of the Galaxy without unnecessary fine details over the entire considered range of distances $R$. An example of a Galactic rotation curve with different numbers of derivatives in the expansion of $\Omega$ is given, for example, in Bobylev et al., 2020) when analyzing galactic masers.

 \subsection{Choosing the value of $R_0$}
 Currently, a number of works have been done on determining the average value of the distance from the Sun to the center of the Galaxy using individual definitions of this quantity,
obtained in the last decade by independent methods. For example,
 $R_0=8.0\pm0.2$~kpc (Vall\'ee, 2017),
 $R_0=8.4\pm0.4$~kpc (de Grijs \& Bono, 2017) or
 $R_0=8.0\pm0.15$~kpc (Camarillo et al., 2018).

 We also note some of the first-class individual definitions of this quantity made recently. An estimate of $R_0=7.9\pm0.3$~kpc (Hirota et al., 2020) was obtained for the masers of the Japanese VERA program. In Gravity Collaboration (2019), from an analysis of a 16-year-long series of observations of the motion of star S2 around a supermassive black hole in the center of the Galaxy, $R_0=8.178\pm0.022$~kpc was found. In Do et al. (2019), based on an independent analysis of the orbit of star S2, $R_0=7.946\pm0.032$~kpc was found. Based on the above results, in the present work, we assume the value $R_0=8.0\pm0.15$~kpc.

\subsection{Distance calculation}\label{Deltapi}
Since the publication of the Gaia DR2 catalog, a problem with trigonometric parallaxes of Gaia DR2 has been known, namely, the need to correct $\Delta\pi$ with a value from $0.03$ to $0.05$ mas (Gaia Collaboration, 2018b,c). Taking into account the results of determining the value of this correction, obtained in Riess et al. (2018), Yalyalieva et al. (2018), Zinn et al. (2019), we add to all the original parallaxes of stars from the Gaia DR2 catalog a correction of $0.05$ ~mas, i.e. $\pi_{new}=\pi+0.05$ ~mas.

\subsection{Residual Velocity Formation}
For formation of the residual velocities, we take into account primarily the peculiar velocity of the Sun, $U_\odot, V_\odot$ and $W_\odot$. It is also necessary to take into account the influence of the differential rotation of the Galaxy. The expressions for accounting for these effects are as follows:
 \begin{equation}
 \begin{array}{lll}
 V_l=V^*_l-[U_\odot\sin l-V_\odot\cos l-r\Omega_0\cos b\\
 +(R-R_0)(R_0\cos l-r\cos b)\Omega'_0
 +0.5(R-R_0)^2(R_0\cos l-r\cos b)\Omega''_0],
 \label{EQU-2}
 \end{array}
 \end{equation}
  \begin{equation}
 \begin{array}{lll}
 V_b=V^*_b-[U_\odot\cos l\sin b + V_\odot\sin l \sin b-W_\odot\cos b \\
 -R_0(R-R_0)\sin l\sin b\Omega'_0
 -0.5R_0(R-R_0)^2\sin l\sin b\Omega''_0],
 \label{EQU-3}
 \end{array}
 \end{equation}
where $V^*_l,V^*_b$~standing on the right-hand sides of the equations are the initial, uncorrected velocities, and on the left-hand sides there are the corrected velocities $V_l,V_b$.

 \subsection{The Residual Velocity Ellipsoid}
We estimated the dispersion of the stellar residual velocities using the following well-known method (Ogorodnikov, 1965). Let $U,V,W$ be the velocities along the coordinate axes $x, y,$ and $z.$ Let us
consider the six second-order moments $a, b, c, f, e, d:$
\begin{equation}
 \begin{array}{lll}
 a=\langle U^2\rangle-\langle U^2_\odot\rangle,\\
 b=\langle V^2\rangle-\langle V^2_\odot\rangle,\\
 c=\langle W^2\rangle-\langle W^2_\odot\rangle,\\
 f=\langle VW\rangle-\langle V_\odot W_\odot\rangle,\\
 e=\langle WU\rangle-\langle W_\odot U_\odot\rangle,\\
 d=\langle UV\rangle-\langle U_\odot V_\odot\rangle,
 \label{moments}
 \end{array}
 \end{equation}
which are the coefficients of the equation for the surface
 \begin{equation}
 ax^2+by^2+cz^2+2fyz+2ezx+2dxy=1,
 \end{equation}
and also the components of the symmetric tensor of moments of the residual velocities:
 \begin{equation}
 \left(\matrix {
  a& d & e\cr
  d& b & f\cr
  e& f & c\cr }\right).
 \label{ff-5}
 \end{equation}
To determine the values in this tensor in the absence of radial-velocity data, the following three equations are used:
\begin{equation}
 \begin{array}{lll}
 V^2_l= a\sin^2 l+b\cos^2 l\sin^2 l-2d\sin l\cos l,
 \label{EQsigm-2}
 \end{array}
 \end{equation}
\begin{equation}
 \begin{array}{lll}
 V^2_b= a\sin^2 b\cos^2 l+b\sin^2 b\sin^2 l+c\cos^2 b\\
 -2f\cos b\sin b\sin l-2e\cos b\sin b\cos l
 +2d\sin l\cos l\sin^2 b,
 \label{EQsigm-3}
 \end{array}
 \end{equation}
\begin{equation}
 \begin{array}{lll}
 V_lV_b= a\sin l\cos l\sin b+b\sin l\cos l\sin b\\
 +f\cos l\cos b-e\sin l\cos b+d(\sin^2 l\sin b-\cos^2\sin b),
 \label{EQsigm-4}
 \end{array}
 \end{equation}
for which least-squares solutions were obtained for the six unknowns $a,b,c,f,e,d$. We then found the eigenvalues of the tensor (\ref{ff-5}): $\lambda_{1,2,3}$ by solving the secular equation:
 \begin{equation}
 \left|\matrix
 {
a-\lambda&          d&        e\cr
       d & b-\lambda &        f\cr
       e &          f&c-\lambda\cr
 }
 \right|=0.
 \label{ff-7}
 \end{equation}
This equation's eigenvalues are inverse squares of the semi-axes of the ellipsoid of the velocity moments; at the same time, they are the squares of the semi-axes of the residual velocity ellipsoid:
 \begin{equation}
 \begin{array}{lll}
 \lambda_1=\sigma^2_1, \lambda_2=\sigma^2_2, \lambda_3=\sigma^2_3,\qquad
 \lambda_1>\lambda_2>\lambda_3.
 \end{array}
 \end{equation}
We found the directions of the main axes of the tensor (\ref{ff-7}), $L_{1,2,3}$ and $B_{1,2,3},$ from the relations:
 \begin{equation}
 \tan L_{1,2,3}={{ef-(c-\lambda)d}\over {(b-\lambda)(c-\lambda)-f^2}},
 \label{ff-41}
 \end{equation}
 \begin{equation}
 \tan B_{1,2,3}={{(b-\lambda)e-df}\over{f^2-(b-\lambda)(c-\lambda)}}\cos L_{1,2,3}.
 \label{ff-42}
 \end{equation}

 \begin{figure*}
  {\begin{center}
 \includegraphics[width=140mm]{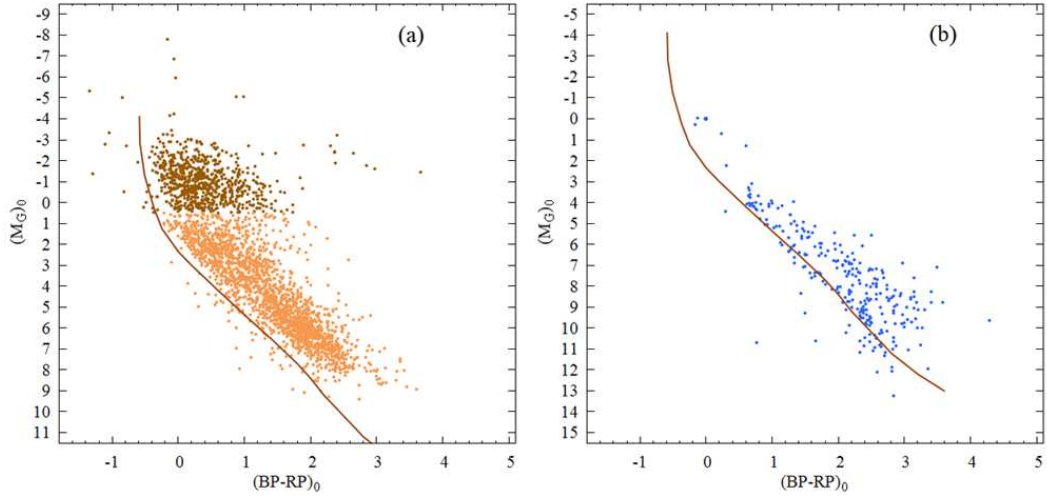}
 \caption{CMD diagram for stars from the catalog of Vioque et al. (2020), with distances greater than 0.5 kpc~(a); with distances less than 0.5 kpc~(b).}
 \label{f-1}
 \end{center}}
 \end{figure*}
 \begin{figure*}
 {\begin{center}
 \includegraphics[width=140mm]{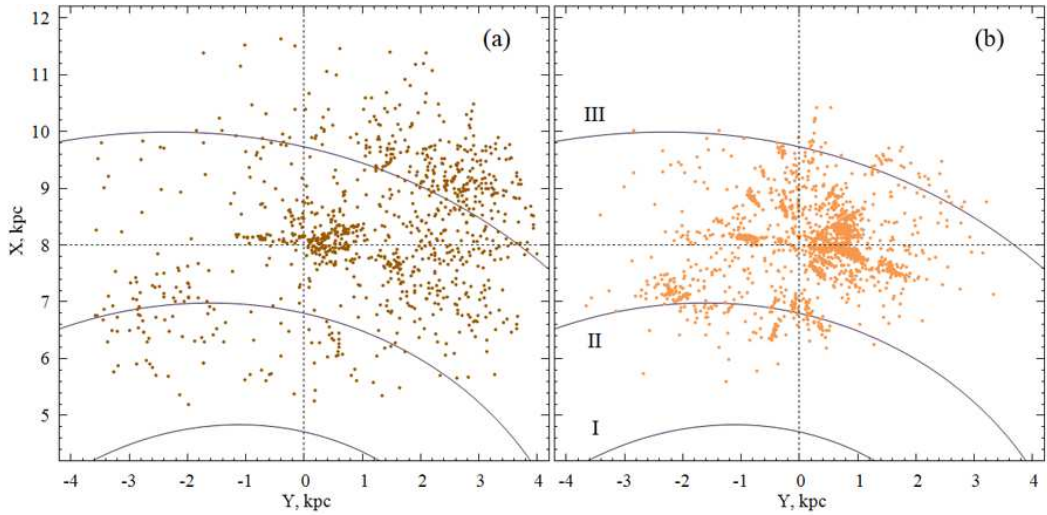}
 \caption{
The distribution of stars from the catalog of Vioque et al. (2020) on the $XY$ plane
with distances greater than 0.5 kpc and with $(M_G)_0\leq0.5^m$~(a);
with distances less than 0.5 kpc and with $(M_G)_0>0.5^m$~(b), the Roman
numerals in the figure number the following spiral arm segments: Scutum (I) , Carina-Sagittarius (II), and Perseus (III).
  }
 \label{f-2}
 \end{center} }
 \end{figure*}

\section{Data}\label{data}
\subsection{Vioque et al. data}
In Vioque et al. (2020), HAeBe stars were selected according to the Gaia DR2 catalog using photometric measurements from infrared surveys 2MASS~(Skrutskie et al., 2006), WISE~(Cutri et al., 2013), IPHAS~(Drew et al., 2005, Barentsen et al., 2014) and VPHAS+ (Drew et al., 2014). The catalog contains three star samples --- PMS, CBe and EITHER.

The PMS sample consists of 8470 candidates for young stars not reached the main sequence. Members of this sample are both Herbig stars Ae/Be, and stars of the T\,Taurus type.

The CBe sample contains 693 candidates for classic Be stars. CBe are stars of spectral class B, located on the main sequence, have fast axial rotation, they are surrounded by gas disks. These are the most massive stars among those considered by us.

The EITHER sample includes 1309 stars. It includes stars that have the following probability values $p$:  $(p_{\rm PMS}+p_{\rm CBe})>50\%,$  but $p_{\rm PMS}<50\%$ and $p_{\rm CBe}<50\%$.

All of these samples contain very young stars. There are no intersections between samples. In order to attract the maximum number of young stars in the problem of determining the rotation parameters of the Galaxy, we also formed a combined sample ALL, which included the stars of the three samples CBe, EITHER and PMS.

The Color-Magnitude Diagram for stars of the ALL sample from the Vioque et al. (2020) catalog with distances greater than 0.5 kpc is given in Fig.~\ref{f-1}(a), and with distances less than 0.5 kpc in Fig.~\ref{f-1}(b).
In Fig.~\ref{f-1}(a) the separation was performed in absolute value with the boundary $(M_G)_0=0.5^m$.

The distribution of stars of this sample with relative trigonometric parallax errors of less than 10\% on the Galactic plane $XY$ is given in Fig.~\ref{f-2}, shows a pattern with a pitch angle of $-13^\circ$ according to Bobylev \& Bajkova (2014). Connection with other segments of the global spiral structure is also visible, although to a more less extent. For example, in the distribution of more massive stars (Fig.~\ref{f-2}(a)), you can see the concentration of stars in the second Galactic quadrant near the Perseus arm. And in the distribution of less massive stars (Fig.~\ref{f-2}(b)) several condensations are clearly visible near the Carina-Sagittarius arm.

\subsection{Marton et al. data}
Marton et al. (2019) obtained a sample of young stars from a combination of Gaia DSR2 and WISE catalogs data using dust transparency indices ($\tau$) from Planck~(Plank Collaboration, 2016). In total, the Marton et al. (2019) catalog contains more than 100~million objects of various nature. For a detailed study of the young population of the Milky Way, they compiled a list of more than 1.1 million candidates for young star objects. These authors introduced 4 main classes --- YSO, extragalactic objects, main sequence stars, and evolved stars. For each object, the probability of belonging to each of the four classes considered is calculated.

In Krisanova et al. (2020), the restrictions for the selection of YSO stars were as follows: line-of-sight velocity error $\leq$10 km s$^{-1}$, heliocentric distance <3 kpc, and relative parallax error <15\%. And according to the probabilities from Marton et al. (2019) it was R>0.5, LY $\geq$0.8, where $R$ is a probability that WISE W3 and W4 detections are real and LY is a probability of the source being a YSO using all WISE bands. And when solving the basic kinematic equations, we got a rather large value of the error of a unit of weight, about 16~km s$^{-1}$.

Experimentally, we found the conditions for the selection of stars, which ultimately have very small variances of their residual velocities. These are the following restrictions on the probability values:
\begin{equation}
  \begin{array}{lll}
 {\rm LY}>0.95~ {\rm and} ~{\rm SY}>0.98~ {\rm and}\\
 {\rm LMS}<0.5,~ ~{\rm SMS}<0.5,~ ~{\rm SE}<0.5~ {\rm and} ~{\rm SEG}<0.5,
 \label{prob}
 \end{array}
\end{equation}
where
  SY is a probability of the source being a YSO without W3 and W4 WISE bands,
  LMS is a probability of the source being a main-sequence star using all WISE bands,
  SMS is a probability of the source being a main-sequence star without W3 and W4 WISE bands,
  SE is a probability of the source being an evolved star without W3 and W4 WISE bands and
  SEG is a probability of the source being an extragalactic object without W3 and W4 WISE bands.

With this selection, we finally have 5571 stars with relative trigonometric parallax errors of less than 10\%, they are located no further than 4 kpc from the Sun.

\begin{figure*}
 {\begin{center}
 \includegraphics[width=140mm]{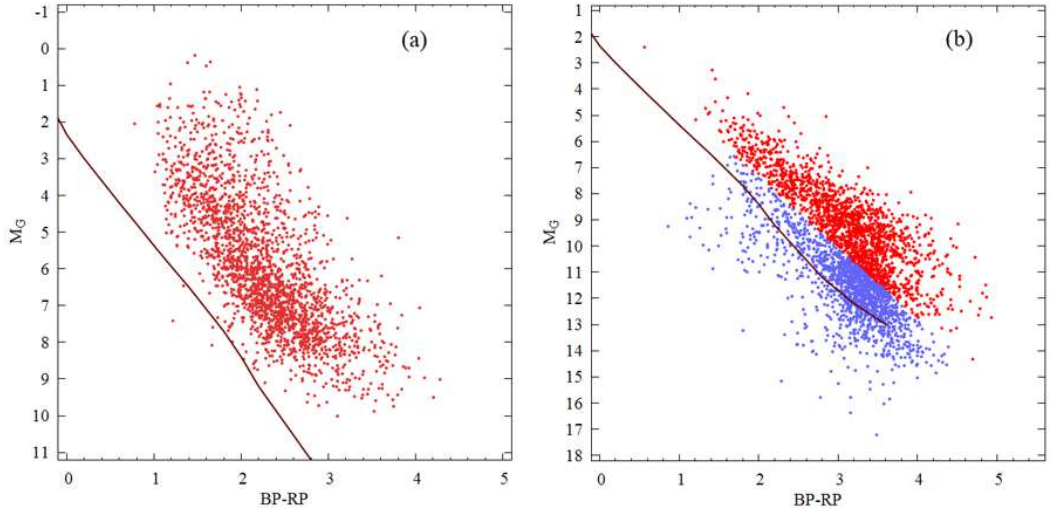}
  \caption{
 CMD diagram for stars from the catalog of Marton et al. (2019) with distances greater than 0.5 kpc~(a);
 with distances less than 0.5 kpc (b). }
 \label{f-DGR-Marton}
 \end{center} }
 \end{figure*}

The CMD diagram for stars from the Marton et al. (2019) catalog with distances greater than 0.5 kpc is given in Fig.~\ref{f-DGR-Marton}(a), and with distances less than 0.5 kpc in Fig.~\ref{f-DGR-Marton}(b). Note that the diagram is constructed without taking into account absorption. As can be seen from Fig.~\ref{f-DGR-Marton}(a), all relatively distant stars in the CMD diagram lie above the main sequence. Unlike the sample from the Vioque et al. (2020) catalog, here we have quite a few stars from the Gould Belt region. Moreover, about half of them, apparently, are old low-mass stars of the main sequence. Therefore, we divided the sample of nearby stars into two parts with the boundary $M_G=2.7(BP-RP)+2^m$.

The division of stars close to the Sun into two parts was made because these two samples have very strong differences both space distribution and kinematic properties.

 \begin{figure*}
 {\begin{center}
 \includegraphics[width=140mm]{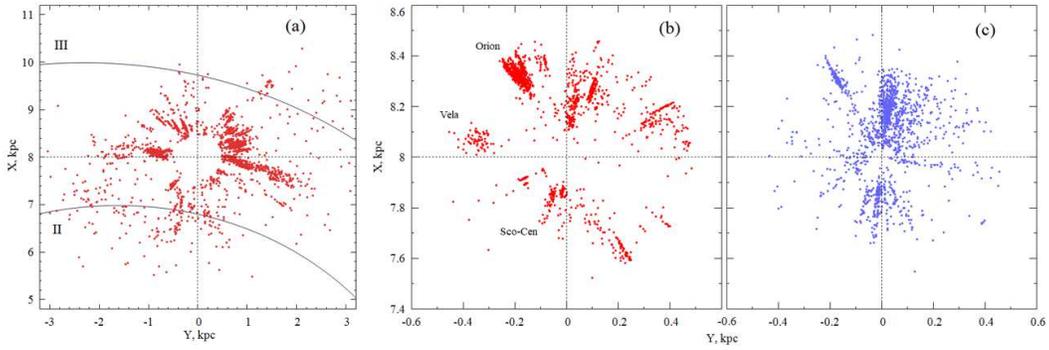}
 \caption{
The distribution of stars from the catalog of Marton et al. (2019) on the $XY$ plane with distances greater than 0.5 kpc~(a);  with distances less than 0.5 kpc~(b) and (c), the Roman
numerals in the figure number the two spiral arm segments: Carina-Sagittarius (II) and Perseus (III).
  }
 \label{f-XY-Marton}
 \end{center} }
 \end{figure*}

The distribution of stars selected by us from the catalog of Marton et al. (2019) with relative trigonometric parallax errors of less than 10\% on the Galactic plane $XY$ is given in Fig.~\ref{f-XY-Marton}. Fig.~\ref{f-XY-Marton}(a) shows a spiral pattern with a pitch angle of $-13^\circ$ according to Bobylev \& Bajkova (2014).
It is interesting to note that the distribution of stars in Fig.~\ref{f-XY-Marton}(a) is very similar to the distribution of Fig.~\ref{f-2}(b), where the connection with the Local spiral arm (Orion arm) is also visible.
The distribution of stars in Fig.~\ref{f-XY-Marton}(b) is very similar to the distribution of stars and OB associations characteristic of the Gould belt. But the distribution of stars in Fig.~\ref{f-XY-Marton}(c) is already much less similar to the distribution of stars in the Gould Belt.

 \begin{table*}
 \caption[]{\small
Kinematic parameters found from the stars of the ALL sample with relative trigonometric parallax errors of less than 10\%, catalog of Vioque et al. (2020).
 }
  \begin{center}  \label{t:01}  \small
  \begin{tabular}{|l|r|r|r|r|r|}\hline
     Parameters & $r>0.5$~kpc &        $r>0.5$~kpc &     $r>0.5$~kpc & $r\leq0.5$~kpc \\
              &         All &  $(M_G)_0\leq0.5^m $ &   $(M_G)_0>0.5^m$ & \\\hline
       $U_\odot,$ km s$^{-1}$ & $6.79\pm0.19$ & $6.49\pm0.47$ & $7.01\pm0.19$ & $12.68\pm0.80$ \\
       $V_\odot,$ km s$^{-1}$ & $9.49\pm0.33$ & $9.25\pm0.80$ & $9.93\pm0.40$ & $15.64\pm1.45$ \\
       $W_\odot,$ km s$^{-1}$ & $7.58\pm0.15$ & $6.65\pm0.29$ & $8.00\pm0.16$ &  $6.24\pm0.52$ \\

 $\Omega_0,$ km s$^{-1}$ kpc$^{-1}$ & $28.60\pm0.13$ & $28.81\pm0.23$ & $28.03\pm0.18$ & $30.8\pm2.5$ \\
 $\Omega'_0,$ km s$^{-1}$ kpc$^{-2}$&$-4.043\pm0.042$&$-4.044\pm0.077$&$-4.012\pm0.050$&$-3.78\pm0.63$\\
 $\Omega''_0,$ km s$^{-1}$ kpc$^{-3}$&$0.746\pm0.035$&$0.701\pm0.056$ & $0.895\pm0.069$ &     --- \\
   $\sigma_0,$ km s$^{-1}$ &         7.80  &         8.98  &          6.91 &       9.82 \\
        $V_0,$ km s$^{-1}$ & $228.8\pm4.4$ & $230.5\pm4.7$ & $224.3\pm4.4$ & $246\pm21$ \\
     $N_\star$       &          2902 &           915 &          1987 &        375 \\
  \hline
 $\sigma_1,$ km s$^{-1}$ & $11.49\pm1.04$ & $13.67\pm2.40$ & $9.45\pm0.94$ & $14.9\pm1.1$\\
 $\sigma_2,$ km s$^{-1}$ & $ 8.89\pm0.83$ & $ 9.25\pm2.44$ & $6.99\pm0.43$ & $10.0\pm1.2$\\
 $\sigma_3,$ km s$^{-1}$ & $ 7.07\pm0.35$ & $ 7.26\pm0.88$ & $6.61\pm0.32$ & $ 7.6\pm1.8$\\

 $L_1, B_1$ & $~86^\circ,$ $12^\circ$ & $~68^\circ,$  $~0^\circ$ & $~79^\circ,$ ~$-17^\circ$ & $~30^\circ,$ $-2^\circ$\\
 $L_2, B_2$ & $176^\circ,$ $-3^\circ$ & $158^\circ,$ ~$~0^\circ$ & $169^\circ,$ ~$~~1^\circ$ & $119^\circ,$ $16^\circ$\\
 $L_3, B_3$ & $280^\circ,$ $78^\circ$ & $296^\circ,$ ~$90^\circ$ & $~82^\circ,$ ~$~73^\circ$ & $306^\circ,$ $74^\circ$\\
  \hline
 \end{tabular}\end{center} \end{table*}
 \begin{table*}
 \caption[]{\small
Kinematic parameters found from stars with relative trigonometric parallax errors of less than 10\%, catalog of Marton et al. (2019).
 }
  \begin{center}  \label{t:02}  \small
  \begin{tabular}{|l|r|r|r|r|r|}\hline
    Parameters & $r:0.5-4$~kpc &        $r\leq0.5$~kpc &     $r\leq0.5$~kpc \\\hline
       $U_\odot,$ km s$^{-1}$ & $7.14\pm0.15$ & $12.48\pm0.25$ & $10.01\pm0.93$ \\
       $V_\odot,$ km s$^{-1}$ & $9.63\pm0.29$ & $14.78\pm0.33$ & $19.10\pm0.91$ \\
       $W_\odot,$ km s$^{-1}$ & $7.74\pm0.12$ & $ 6.86\pm0.14$ & $ 7.25\pm0.50$ \\

    $\Omega_0,$ km s$^{-1}$ kpc$^{-1}$ &  $28.40\pm0.14$  & $28.23\pm0.98$ & $36\pm5$\\
   $\Omega'_0,$ km s$^{-1}$ kpc$^{-2}$ & $-3.961\pm0.041$ &$-2.50\pm0.19$ &$-4.012\pm0.050$\\
  $\Omega''_0,$ km s$^{-1}$ kpc$^{-3}$ &  $0.859\pm0.060$ &           --- &      --- \\

  $\sigma_0,$ km s$^{-1}$ &         5.55  &     5.05  &      18.13 \\
      $V_0,$ km s$^{-1}$  & $227.2\pm4.4$ & $226\pm9$ & $287\pm42$ \\
       $N_\star$   &          2277 &      1748 &       1536 \\\hline

 $\sigma_1,$ km s$^{-1}$ & $6.91\pm0.41$ & $9.96\pm0.14$ & $25.50\pm0.17$\\
 $\sigma_2,$ km s$^{-1}$ & $6.11\pm0.20$ & $5.40\pm0.96$ & $17.40\pm1.65$\\
 $\sigma_3,$ km s$^{-1}$ & $5.09\pm0.22$ & $ 0.9\pm3.8 $ & $13.91\pm0.61$\\

 $L_1, B_1$&$~15^\circ,$ $~9^\circ$ &$~64^\circ,$ ~$21^\circ$ & $~33^\circ,$ ~$-3^\circ$ \\
 $L_2, B_2$&$106^\circ,$ $10^\circ$ &$153^\circ,$ ~$~4^\circ$ & $121^\circ,$ ~$31^\circ$ \\
 $L_3, B_3$&$243^\circ,$ $77^\circ$ &$256^\circ,$ ~$68^\circ$ & $308^\circ,$ ~$59^\circ$ \\
  \hline
 \end{tabular}\end{center} \end{table*}

\section{RESULTS AND DISCUSSION}
Table \ref{t:01} gives the values of kinematic parameters found from the stars from the catalog of Vioque et al. (2020). The data are divided into distant $r>0.5$~kpc, which are given in the first three columns and nearby stars $r\leq0.5$, presented in the last column. In the second and third columns, the separation was performed in absolute value with the boundary $(M_G)_0=0.5^m$, as in Fig.~\ref{f-1}, and in the first column of the Table, the results were obtained from a combined sample of distant stars. As can be seen from Fig.~\ref{f-1}(b), there are few nearby stars from the catalog of Vioque et al. (2020), they are on almost 50\% contaminated with main sequence stars. The kinematics of young close stars has features associated with their belonging to the Gould Belt. Therefore, when searching for Galactic rotation parameters, it is better not to use them.

Table \ref{t:02} gives the values of the kinematic parameters found from the stars from the catalog of Marton et al. (2019). The data are divided into distant $r>0.5$~kpc, which are given in the first column and nearby stars with $r\leq0.5$ kpc, presented in the second and third columns. The separation of nearby stars was carried out in absolute value relative to the empirical dependence $M_G=2.7(BP-RP)+2^m$, as in Fig.~\ref{f-DGR-Marton}. As can be seen from Fig.~\ref{f-DGR-Marton}(b) and Fig.~\ref{f-DGR-Marton}(c), as well as the Table \ref{t:02}, the two samples of nearby stars have strong differences, both in the distribution on the $XY$ plane and in kinematics.

\subsection{Galaxy Rotation}
As can be seen from the tables \ref{t:01} and \ref{t:02} for all samples of stars with distances greater than 0.5 kpc, the rotation parameters of the Galaxy $\Omega_0,$ $\Omega'_0$ and $\Omega''_0$ are well defined. These values are in good agreement both among themselves and with the results of the analysis of other samples of young objects.

In order to reduce random errors of the determined parameters of the Galactic rotation when solving the system of equations (\ref{EQ-2})--(\ref{EQ-3}), we formed a combined sample of distant stars. Taking into account the fact that between the samples of distant stars from the catalogs of Vioque et al. (2020) and Marton et al. (2019) there are about 700 common stars, a sample of 4431 stars was obtained. From this sample of distant ($r:0.5-4$~kpc) stars, there were found the components of the group velocity vector $(U_\odot,V_\odot,W_\odot)=(7.06,9.16,7.61)\pm(0.14,0.24,0.11)$~km s$^{-1}$
and the following components of the angular velocity of Galactic rotation:
 \begin{equation}
 \label{sol-common}
 \begin{array}{lll}
  \Omega_0=~28.63\pm0.10~\hbox{km s$^{-1}$ kpc$^{-1}$},\\
  \Omega^{\prime}_0=-4.007\pm0.032~\hbox{km s$^{-1}$ kpc$^{-2}$},\\
 \Omega^{\prime\prime}_0=~0.710\pm0.028~\hbox{km s$^{-1}$ kpc$^{-3}$}.
 \end{array}
 \end{equation}
In this solution, the unit weight error is $\sigma_0=7.2$~km s$^{-1}$, the values of the Oort constants
  $A=16.03\pm0.33$~km s$^{-1}$ kpc$^{-1}$ and $B=-12.60\pm0.34$~km s$^{-1}$ kpc$^{-1}$, and the linear
velocity of rotation of the Solar neighborhood around the center of the Galaxy is $V_0=229.1\pm4.4$~km s$^{-1}$. It can be seen that in the solution (\ref{sol-common}), random errors of the determined parameters decreased in comparison with the results specified in Tables \ref{t:01} and \ref{t:02}.

Based on 130 masers with measured VLBI trigonometric parallaxes, Rastorguev et al. (2017) found the following parameters of the Galactic rotation curve:
 $\Omega_0=28.93\pm0.53$ km s$^{-1}$ kpc$^{-1}$,
 $\Omega'_0=-3.96\pm0.07$ km s$^{-1}$ kpc$^{-2}$,
 $\Omega''_0=0.87\pm0.03$ km s$^{-1}$ kpc$^{-3}$, and
  $V_0=243\pm10$ km s$^{-1}$ (for $R_0=8.40\pm0.12$ kpc found).

In work Bobylev \& Bajkova (2019b) we have studied a sample containing $\sim6000$ OB stars with proper motions and trigonometric parallaxes from the Gaia DR2 catalogue. The following parameters of the angular velocity of Galactic rotation have been found:
 $\Omega_0=29.70\pm0.11$ km s$^{-1}$ kpc$^{-1}$,
 $\Omega'_0=-4.035\pm0.031$ km s$^{-1}$ kpc$^{-2}$, and
 $\Omega''_0= 0.620\pm0.014$ km s$^{-1}$ kpc$^{-3}$. The circular rotation velocity of the solar neighborhood around the Galactic center is (for the adopted $R_0=8.0\pm0.15$ kpc).

 It is interesting to note the work of Eilers et al. (2019), where, for a sample of more than 25\,000 luminous red giant stars, the estimate $V_0=229.0\pm0.2$ km s$^{-1}$ (for the adopted $R_0=8.122\pm0.031$~kpc) was obtained. Rotation curve of the Galaxy was constructed on the distance interval $R:5-25$~kpc.

 In a recent work by Ablimit et al. (2020), about 3\,500 classical Cepheids from various sources were used to construct the rotation curve of the Galaxy. The circular rotation velocity of the solar neighbourhood was $V_0=232.5\pm0.9$~km s$^{-1}$, which is in good agreement with our estimate of this speed. Based on the Cepheids of this sample, the rotation curve of the Galaxy was constructed on the distance interval $R:4-19$~kpc.

 \begin{figure*}
 {\begin{center}
 \includegraphics[width=150mm]{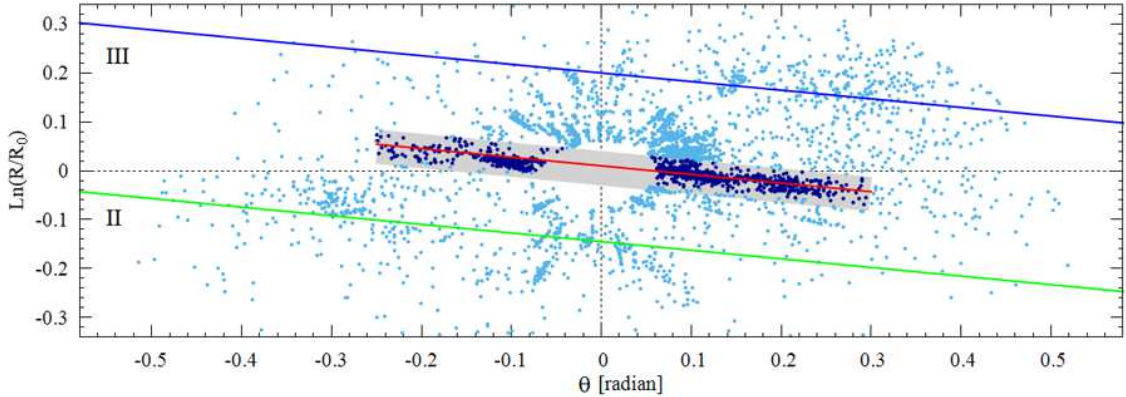}
 \caption{
The distribution of the combined sample of 4431 distant stars on the plane $\ln (R/R_0)-\theta$, the Roman
numerals in the figure number the Carina-Sagittarius (II) and Perseus (III) spiral arm segments.
  }
 \label{f-tani}
 \end{center} }
 \end{figure*}

\subsection{Residual Velocity}
Bobylev (2020) performed a kinematic analysis of stars of the type T\,Tau from the list of Zari et al. (2018). The error value of the unit of weight $\sigma_0$ for various samples is in the range of 10--12~km s$^{-1}$, and the values of the Oort constants $A$ and $B$ are close to ones inherent  for the Gould Belt, i.e. significantly different from the characteristics of Galactic rotation. An analysis of the movements of the stars of this sample of stars showed that the residual velocity ellipsoid with the main semiaxes  $\sigma_{1,2,3}=(8.87,5.58,3.03)\pm(0.10,0.20,0.04)$~km s$^{-1}$
is located at an angle of $22\pm1^\circ$ to the Galactic plane with the longitude of the ascending node $298\pm2^\circ$.

The components of the peculiar velocity of the Sun $(U_\odot,V_\odot,W_\odot)$=(12.48, 14.78, 6.86) km s$^{-1}$, the variance of the residual velocities and the orientation parameters of the ellipsoid indicated in the penultimate column of the Table~\ref{t:02} are in good agreement with the estimates characteristic for the Gould Belt. Fig.~\ref{f-DGR-Marton}(b) also shows the close relationship of the stars of this sample with the Gould Belt.

It is also obvious that the velocity value $V_\odot=19.10\pm0.91$ km s$^{-1}$ and the variance of the residual velocities indicated in the last column of Table~\ref{t:02} indicate the dominance of old stars in this sample.

One of the goals of the present work was to select such conditions that would ensure obtaining a sample from the list of Marton et al. (2019) with a minimum dispersion of the residual velocities of stars. The results indicated in the first and second columns of Table~\ref{t:02} indicate that such a goal has been achieved. Of particular interest is the sample which analysis results are indicated in the first column of Table \ref{t:02}. After all, we only analyzed the components $V_l$ and $V_b$, calculated through parallaxes and proper motions. Moreover, random errors of these velocities increase with increasing distance. However, as can be seen from the first column of Table \ref{t:02}, the error of the unit weight $\sigma_0$ and the variance of the residual velocities $\sigma_{1,2,3}$ of these stars are small. Thus, the application of constraints (\ref{prob}) in combination with the condition $r>0.5$~kpc (if necessary, with a constraint on $\sigma_\pi/\pi$) allows one to select a uniform sample of very young stars from the catalog of Marton et al. (2019).

The value $\sigma_0$ can be considered as the average coordinate dispersion of residual velocities over three coordinates. The average dispersion of hydrogen clouds HI is 3--5~km s$^{-1}$ (Clemens, 1985). The dispersion of the residual velocities of distant OB stars is 10--12~km s$^{-1}$ (Uemura et al., 2000, Bobylev \& Bajkova 2019b). For example, in the kinematic analysis of $\sim6000$ OB stars with proper motions and trigonometric parallaxes from the Gaia DR2 catalog, the unit error was $\sigma_0=11$~km s$^{-1}$ (Bobylev \& Bajkova, 2019b). Thus, the found value of $\sigma_0\sim6$~km s$^{-1}$ indicates the extreme youth of the analyzed stars recently formed from hydrogen clouds.

In the catalog of Vioque et al. (2020), almost all stars are very young. According to their positions on the CMD diagram, they are generally more massive compared to the stars of our sample from the catalog of Marton et al. (2019). As can be seen from Table \ref{t:01}, the variances of their residual velocities are also small. The third axis of the ellipsoid of residual velocities found from distant stars does not deviate from the vertical, which is clearly seen in the sample from the second column of Table \ref{t:01}, where $B_3=90^\circ$.

\subsection{Local Arm}
The equation describing the position of a Galactic object on the logarithmic spiral can be written in the following way:
 \begin{equation}
 R=a_0 e^{(\theta-\theta_0)\tan i},
 \label{spiral-1}
 \end{equation}
where $a_0>0$, $\theta$~is the object's position angle measured in the direction of Galactic rotation: $\tan\theta = y/(R_0-x)$, where $x,y$~are Galactic heliocentric rectangular coordinates of
the object; $\theta_0$ is angle at which $R=a_0$; $i$ is pitch angle. 

Putting $a_0=R_0$ in Equation~(\ref{spiral-1}), we can estimate the value of pitch angle $i$ as
\begin{equation}
  \tan i=\frac{\ln (R/R_0)}{\theta-\theta_0},
 \label{spiral-04}
\end{equation}
where, obviously, $\theta_0=0^\circ$. For this purpose, a ``position angle~-- distance logarithm'' diagram is constructed, where arms of a logarithmic spiral are presented as line segments. Such method is widely used for studying the Galactic spiral structure based on the various object data (Popova \& Loktin, 2005, Xu et al., 2013, Bobylev \& Bajkova, 2014). The advantage of this approach is that the estimate of pitch angle $i$ does not depend on the number of spiral arms.

As can be seen from Fig.~\ref{f-2} and Fig.~\ref{f-XY-Marton}, there are too few stars near the Perseus and Carina-Sagittarius spiral arms  to specify the value of the pitch angle $i$. But the section of the Local arm is very well visible. Therefore, we decided to find the value of the pitch angle $i$ from the combined sample of 4431 distant stars. In Fig.~\ref{f-tani} the distribution of the combined sample of 4431 distant stars on the plane $\ln (R/R_0)-\theta$ is given.

The spiral arms characteristics were obtained from linear regression $\ln(R/R_0)=a\cdot\theta+b$ (see eq, (\ref{spiral-04})). The problem was solved both with unit weights.
Using 1212 stars, candidates for belonging to the Local arm, the following parameters were found
$a=-0.1562\pm0.0023$ and $b=0.0075\pm0.0003$, which implies that $i=-8.9\pm0.1^\circ$ and the Local arm near the Sun extends toward the anticenter of the Galaxy at a distance $60\pm2$~pc ($x=-60\pm2$~pc).

The selection of stars for such an analysis was carried out by us in the area marked by a gray stripe in Fig.~\ref{f-tani}. The width of this area is chosen according to the estimate from Reid et al. (2019). All stars that fall into this area are marked in the figure with dark blue dots. The red line in the Figure shows a segment of the Local Arm with the found parameters, Carina-Sagittarius and Perseus spiral arm segments with the same pitch angle are also given.

Xu et al. (2013) obtained an estimate of the pitch angle of the Local Arm $i=-10.1\pm2.7^\circ$ using 30 masers with VLBI-measured parallaxes.  With increasing number of measurements, these authors found $i=-11.6\pm1.8^\circ$ (Xu et al., 2016). In work of Reid et al. (2019), a new estimate of $i=-11.4\pm1.9^\circ$ was obtained from 28 sources.

\section{CONCLUSIONS}
The kinematic properties of young stars that have not yet reached the stage of the main sequence are studied. The selection of these stars was carried out in the works of Marton et al. (2019) and Vioque et al. (2020) according to the Gaia DR2 catalog and using various photometric infrared surveys. The approaches of these authors have methodological differences, therefore, the properties of their samples have both differences and similarities. We examined stars with relative trigonometric parallax errors of less than 10\%.

First of all, it can be noted that in the samples from work of Vioque et al. (2020), all the stars are very young. Among them, the percentage of old main sequence stars is negligible. In addition, among them there are few stars from a near-solar neighborhood with a radius of 0.5~kpc.

The catalog of Marton et al. (2019) contains a huge number of candidates in the YSO with different values of the probability of belonging to young objects. Experimentally, we found the conditions for the selection of stars, which ultimately have very small variances of their residual velocities. These are the following restrictions on the probability values: LY>0.95 and SY>0.98, provided LMS<0.5, SMS<0.5, SE<0.5, and SEG<0.5. With this selection, it turns out that all relatively distant stars ($r>0.5$~kpc) in the CMD diagram lie above the main sequence.

From distant stars ($r>0.5$~kpc) from both catalogs, the Galactic rotation curve parameters were determined in good agreement with each other. The linear velocity of the Galactic rotation $V_0$ is in the range 227--229~km s$^{-1}$, and the error in determining this value in both cases is 4.4~km s$^{-1}$. It can be noted that the sample of stars from the catalog of Vioque et al. (2020) contains generally more massive stars compared to the sample from the catalog of Marton et al. (2019). Although in the region of faint stars between these catalogs there are about 15\% of the total stars. The third axis of the ellipsoid of residual velocities found from distant stars does not deviate from the vertical, what is especially clearly seen in the sample of massive stars.

From 4431 proper motions of the combined sample of stars with relative trigonometric parallax errors of less than 10\% and heliocentric distances from 0.5~kpc to 4~kpc, the following values of the angular velocity of the Galaxy were found:
$\Omega_0=28.63\pm0.10$~km s$^{-1}$ kpc$^{-1}$,
 $\Omega^{'}_0=-4.007\pm0.032$~km s$^{-1}$ kpc$^{-2}$ and
 $\Omega^{''}_0=0.710\pm0.028$~km s$^{-1}$ kpc$^{-3}$,
 where the linear velocity of the circular rotation of a solar neighborhood around the center of the Galaxy is $V_0=229.1\pm4.4$~km s$^{-1}$ for the accepted value $R_0=8.0\pm0.15$~kpc.

The variance of the residual velocities of the stars from the list of Vioque et al. (2020) slightly depends on the position of the stars in the CMD diagram. From the stars from the bottom of the diagram with the absolute value $(M_G)_0>0.5^m,$ the following parameters of the ellipsoid of their residual velocities were found: $\sigma_{1,2,3}=(9.45,6.99,6.61)\pm(0.94,0.43,0.32)$~km s$^{-1}$.
And for the stars from the top of the diagram, with the absolute value $(M_G)_0\leq0.5^m,$ their values turned out to be somewhat larger: $\sigma_{1,2,3}=(13.67,9.25,7.26)\pm(2.40,2.44,0.88)$~km s$^{-1}$.

The variance of the residual velocities of stars extracted from the catalog of Marton et al. (2019) is less dependent on the position of the stars in the CMD diagram. They depend on the heliocentric distance of the stars of the sample. Based on the stars from the range of distances $r: 0.5-4$~kpc, the following parameters of the ellipsoid of their residual velocities were found:
  $\sigma_{1,2,3}=(6.91,6.11,5.09)\pm(0.41,0.20,0.22)$~km s$^{-1}$.
And for nearby stars from the vicinity of $r\leq0.5$~kpc we found
  $\sigma_{1,2,3}=(9.96,5.40,0.9)\pm(0.14,0.96,3.8)$~km s$^{-1}$.

Among the stars from the catalog of Marton et al. (2019), lying in a near-solar neighborhood with a radius of 0.5~kpc, the proportion of old main sequence stars is large. Using them, the following parameters of the ellipsoid of their residual velocities were found: $\sigma_{1,2,3}=(25.5,17.4,13.9)\pm(0.2,1.7,0.6)$~km s$^{-1}$.

Distant stars from both catalogs trace the Local spiral arm (Orion arm) well. For 1212 stars of the combined sample, a new estimate of the pitch angle of the Local spiral arm, equal to
$i=-8.9\pm0.1^\circ$ is obtained.


\bigskip
 {\bf REFERENCES}
 {\small

Ablimit I., Zhao G., Flynn C., and Bird S.A., 2020, ApJ {\bf 895L}, 12

Barentsen G., Farnhill H.J., Drew J.E, et al., 2014, MNRAS {\bf 444}, 3230 

Bertout C., Robichon N., and Arenou F., 1999, A\&A {\bf 352}, 574

Bobylev V.V. and Bajkova A.T., 2014, MNRAS {\bf 437}, 1549 

Bobylev V.V. and Bajkova A.T., 2018, Astron. Lett. {\bf 44}, 675 

Bobylev V.V. and Bajkova A.T., 2019a, Astron. Lett. {\bf 45}, 208 

Bobylev V.V. and Bajkova A.T., 2019b, Astron. Lett. {\bf 45}, 331 

Bobylev V.V., 2020, Astron. Lett. {\bf 46}, 131

Bobylev V.V., Krisanova O.I., and Bajkova A.T., 2020, Astron. Lett. {\bf 46}, (in press) 

Camarillo T., Varun M., Tyler M., and Bharat R., 2018, PASP {\bf 130}, 4101

Cantat-Gaudin T., Jordi C., Wright N.J., et al., 2019, A\&A {\bf 626}, 17  

Clemens D.P., 1985, ApJ {\bf 295}, 422

Cutri R.M., Wright E.L., Conrow T.,  
et al., 2013, VizieR On-line Data Catalog: II/311 

Damiani F., Prisinzano L., Pillitteri I., et al., 2019, A\&A {\bf 623}, 112

Do T., Hees A., Ghez A., et al., 2019, Science {\bf 365}, 664

Drew J.E., Greimel R., Irwin M.J., et al., 2005, MNRAS {\bf 362}, 753 

Drew J.E., Gonzalez-Solares E., Greimel R., et al., 2014, MNRAS {\bf 440}, 2036 

Eilers A.-C., Hogg D.W., Rix H.-W., and Ness M.K., 2019, ApJ {\bf 871}, 120

Hirota T., Nagayama T., Honma M., et al., 2020, PASJ {\bf 72}, 50

Gaia Collaboration, F. Arenou, X. Luri, C. Babusiaux, et al., 2018c, A\&A {\bf 616}, 17

Gaia Collaboration, Prusti T., de Bruijne J.H.J., Brown A.G.A., et al., 2016, A\&A {\bf 595},~1 

Gaia Collaboration, Brown A.G.A., Vallenari A., Prusti T., et al., 2018a, A\&A {\bf 616}, 1 

Gaia Collaboration, L. Lindegren, J. Hernandez, A. Bombrun, et al., 2018b, A\&A {\bf 616},~2 

Gravity Collaboration, Abuter R., Amorim A., Baub\"ock N., et al., 2019, A\&A {\bf 625}, L10

de Grijs R., and Bono G., 2017, ApJS {\bf 232}, 22

Grosschedl J.E., Alves J., Stefan Meingast S., et al., 2018, A\&A {\bf 619}, 106

Krisanova O.I., Bobylev V.V., and Bajkova A.T.,  2020, Astron. Lett. {\bf 46}, 370

Mamajek E.E., Meyer M.R., and Liebert J., 2002, AJ {\bf 124}, 1670

Marton G., \'Abrah\'am P., Szegedi-Elek E., et al., 2019, MNRAS {\bf 487}, 2522

Mr\'oz P., Udalski A., Skowron D.M., et al., 2019, ApJ {\bf 870}L, 10

Ogorodnikov K.F., 1965, {\it Dynamics of stellar systems} (Oxford: Pergamon, ed. Beer, A. 1965).

Popova M.E., and Loktin A.V., 2005, Astron. Lett. {\bf 31}, 171

Planck Collaboration, Adam R., Ade P.A.R., Aghanim N., et al., 2016, A\&A {\bf 594}, 10 

Preibisch T., and Zinnecker H., 1999, AJ, {\bf 117}, 2381

Rastorguev A.S., Zabolotskikh M.V., Dambis A.K., et al., 2017, Astrophys. Bull. {\bf 72}, 122

Reid M.J., Menten K.M., Brunthaler A., et al., 2019 ApJ {\bf 885}, 131

Riess A.G., Casertano S., Yuan W., et al., 2018,  ApJ {\bf 861}, 126

Sartori M.J., L\'epine J.R.D., and Dias W.S., 2003, A\&A {\bf 404}, 913

Skrutskie R.M., Cutri R.M., Stiening R., et al., 2006, AJ {\bf 131}, 1163 

Soubiran C., Cantat-Gaudin T., Romero-G\'omez M., et al., 2018, A\&A {\bf 619}, 155

Uemura M., Ohashi H., Hayakawa T., et al., 2000, PASJ {\bf 52}, 143

Vall\'ee J.P., 2017, Ap\&SS {\bf 362}, 79

Vioque M., Oudmaijer R.D., Schreiner M., et al., 2020, A\&A {\bf 638}, 21

Wichmann R., Bastian U., J. Krautter J., et al., 1998, MNRAS {\bf 301}, 39

Xu Y., Li J.J., Reid M.J., et al., 2013, ApJ {\bf 769}, 15

Xu Y., Reid M.J., Dame T., et al., 2016, Sci. Adv. 9, e1600878 

Xu Y., Li J.J., Reid M.J., et al., 2018, A\&A {\bf 616}, 15 

Yalyalieva L.N., Chemel A.A., Glushkova E.V., et al., 2018, Astrophys. Bull. {\bf 73}, 335

Zari E., Hashemi H., Brown A.G.A., et al., 2018, A\&A {\bf 620}, 172

Zinn J.C., Pinsonneault M.H., Huber D., and Stello D., 2019, ApJ {\bf 878}, 136

}
\end{document}